# Single-Shot Simultaneous Intensity, Phase, and Polarization Imaging with Metasurface

Yanjun Bao*, and Baojun Li*

Guangdong Provincial Key Laboratory of Nanophotonic Manipulation, Institute of Nanophotonics, College of Physics & Optoelectronic Engineering, Jinan University, Guangzhou 511443, China

*Corresponding Authors: Y. Bao (yanjunbao@jnu.edu.cn), B. Li (baojunli@jnu.edu.cn)

**Abstract**:Optical imaging of the intensity, phase and polarization distributions of optical field is fundamental to numerous applications. Traditional methods rely on bulky optical components and require multiple measurements. Recently, metasurface-based (MS-based) imaging strategies have emerged as a promising solution to address these challenges. However, they have been primarily limited to capturing partial information of the three parameters, tailored to specific optical fields, which poses challenges when addressing with arbitrary field distributions and achieving three-parameter imaging. In this study, we introduce a MS-based approach for single-shot optical imaging that simultaneously captures all the three parameters of optical fields with arbitrary intensity, phase, and polarization distributions. We experimentally validate the versatility of our method by conducting imaging of various types of optical fields with arbitrary well-defined distributions. The strategy presented in our work is expected to open up promising avenues for diverse applications, including imaging, optical communications, and beyond.

**Keywords**

**INTRODUCTION**

Optical fields are primarily characterized by three parameters: intensity, phase, and polarization. The precise imaging of these parameters holds significant importance across a broad spectrum of applications, including microscopy, material characterization, biophotonics and more. Traditional imaging techniques rely on bulky optical components, such as lenses, waveplates, polarizers, phase retarders, and often require multiple measurements. For instance, lenses can image the intensity in a single shot, while polarization imaging demands several specific rotational combinations of polarizers and waveplates[1]. For phase imaging, techniques such as phase-shifting interferometry[2-5] and transport of intensity equation (TIE)[6-8] are commonly used but also involve multiple measurements. The former requires a minimum of three phase-shifted reference field for interference[2]. The TIE method is susceptible to noise and requires multiple measurements across several planes with varying transmission distances[6]. Additionally, the algorithm of TIE may not converge and result in inaccurate results for specific phase distributions like those with a vortex pattern[6].

Metasurfaces, consisting of two-dimensional planar structures with artificial atoms, have emerged as a pivotal tool for manipulating the intensity, phase, and polarization of light, unlocking advancements in a variety of optical applications[9-20]. Their planar configuration, combined with versatile optical control, presents innovative opportunities for device miniaturization and enabling single-shot imaging. For instance, metasurface-based (MS-based) metalenses[21-28] can replace traditional lenses for intensity imaging, offering significantly reduced thickness and simplified integration. For polarization imaging, metasurfaces has been utilized for single-shot imaging of all four Stokes components[28, 29]. Regarding phase imaging, several methods, such as differential phase imaging (DPI)[30-34] and TIE[35], have been employed with metasurface. However, these approaches impose certain constraints on the input fields. The DPI method relies on the gradient of the optical field, necessitating input fields with uniform intensity[30, 33] to avoid amplitude differentiation effects or linear polarization to ensure two identical polarization components during interference[30-33]. Moreover, such DPI method only captures the gradient along one direction in a single-shot[30, 31, 33], precluding the complete reconstruction of phase distribution. Recently, a strategy combining metasurfaces with polarization cameras has been developed to enable complex amplitude field imaging[34]. For the TIE method, metasurfaces facilitate simultaneous recording of two images at different propagation distances[35], which also requires 45° linear polarization of the incident field to ensure identical TM and TE polarization distributions. Consequently, existing metasurface-based imaging techniques are restricted to capturing partial optical parameters for specific field distributions. Hence, developing a method capable of imaging arbitrary optical fields and achieving single-shot three-parameter imaging is highly desirable.

In this work, we have demonstrated MS-based single-shot imaging of all the three parameters simultaneously for optical fields with arbitrary intensity, phase, and polarization distributions. Our work offers three significant advancements compared to previous MS-based imaging strategies: it addresses arbitrary optical fields, captures all three parameters simultaneously, and achieves single-shot three-parameter imaging without multiple measurements of each parameter. This achievement is rooted in the careful design of different components of the Jones matrix of the metasurface, optimized to diffract the orthogonal polarization of the input field and generate the desired reference fields. We have verified the versatility of our method by conducting optical imaging on various types of arbitrary optical fields with well-defined distributions.

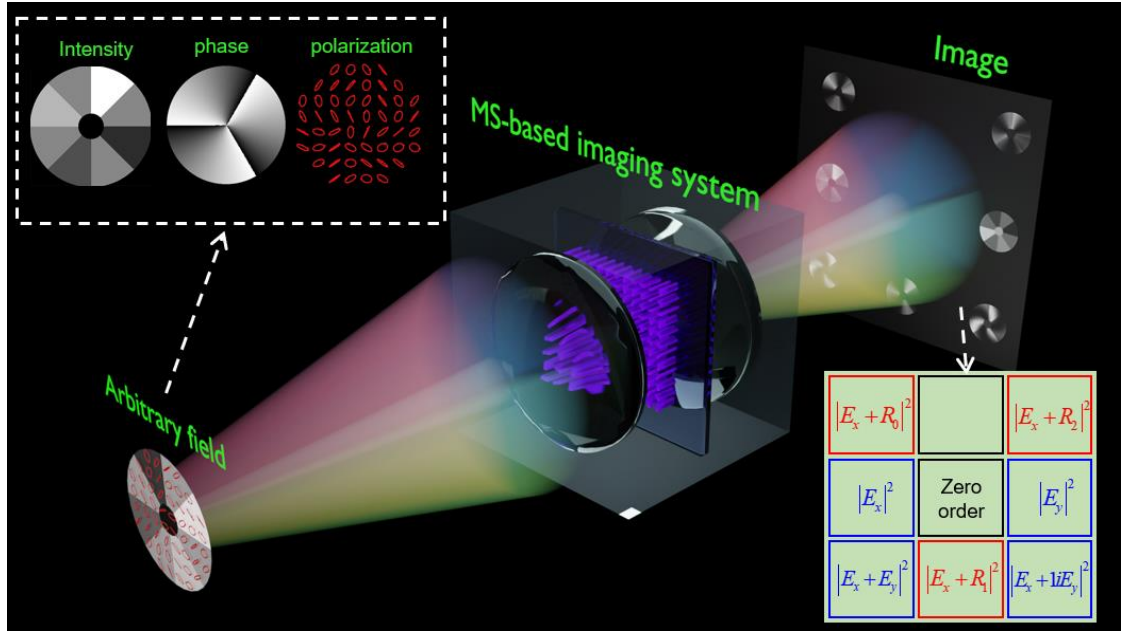

**Figure 1**. MS-based single-shot three-parameter imaging of arbitrary optical field. The input field exhibits arbitrary intensity, phase, and polarization distributions. The captured image contains seven sub-images at distinct regions. Three sub-images (red rectangles) represent the interference between $E_x$ and reference fields $R_m$. The other four sub-images (blue rectangles) capture intensities of $E_x$, $E_y$, $E_x+1iE_y$, and $E_x+E_y$.

## RESULTS AND DISCUSSION

**Design of MS-based single-shot three-parameter imaging**

Figure 1 presents our MS-based optical imaging system, designed for single-shot three-parameter imaging of input field with arbitrary intensity, phase, and polarization distributions. Upon passing through the imaging system, the input field is diffracted to produce seven sub-images located at distinct regions (inset of Figure 1). Among them, three are intentionally designed to represent the interferences between the $x$-component of the electric field ($E_x$) and the reference fields ($R_m$), as highlighted with red rectangles. Here, $R_m$ represents an optical field with uniform amplitude and constant phase $2m\pi/3$, where $m$ takes values 0, 1 and 2. The intensities of the three captured sub-images (labeled $I_0$, $I_1$ and $I_2$) allow us to obtain the $E_x$ phase distributions $\varphi_x$ as follows[2] (see details in Supplementary Section 1)

$$\varphi_x = \mathrm{atan2}\left(\sqrt{3}(I_1 - I_2), 2I_0 - I_1 - I_2\right) \quad (1)$$

where the function atan2($y$, $x$) return the phase of a complex number in the form $x+1\mathrm{i}y$.

The remaining four sub-images, marked with blue rectangles, are designed to capture the intensities of $E_x$, $E_y$, $E_x+1\mathrm{i}E_y$, and $E_x+E_y$, labeled as $I'_0$, $I'_1$, $I'_2$ and $I'_3$, respectively. This procedure is essentially the same as the conventional measurement of the four Stokes polarization parameters[1]. Through these image intensities, we extract not only the intensities of $E_x$ and $E_y$, but also the phase difference between them, which can be calculated as (see details in Supplementary Section 1):

$$\varphi_x - \varphi_y = \mathrm{atan2}\left(I'_2 - I'_0 - I'_1, I'_3 - I'_0 - I'_1\right) \quad (2)$$

In conjunction with Equation (1), we can deduce both the intensity and phase distributions of $E_x$ and $E_y$, which are equivalent to the intensity, phase and polarization of the field. Therefore, such system enables a single-shot three-parameter imaging of arbitrary input fields.

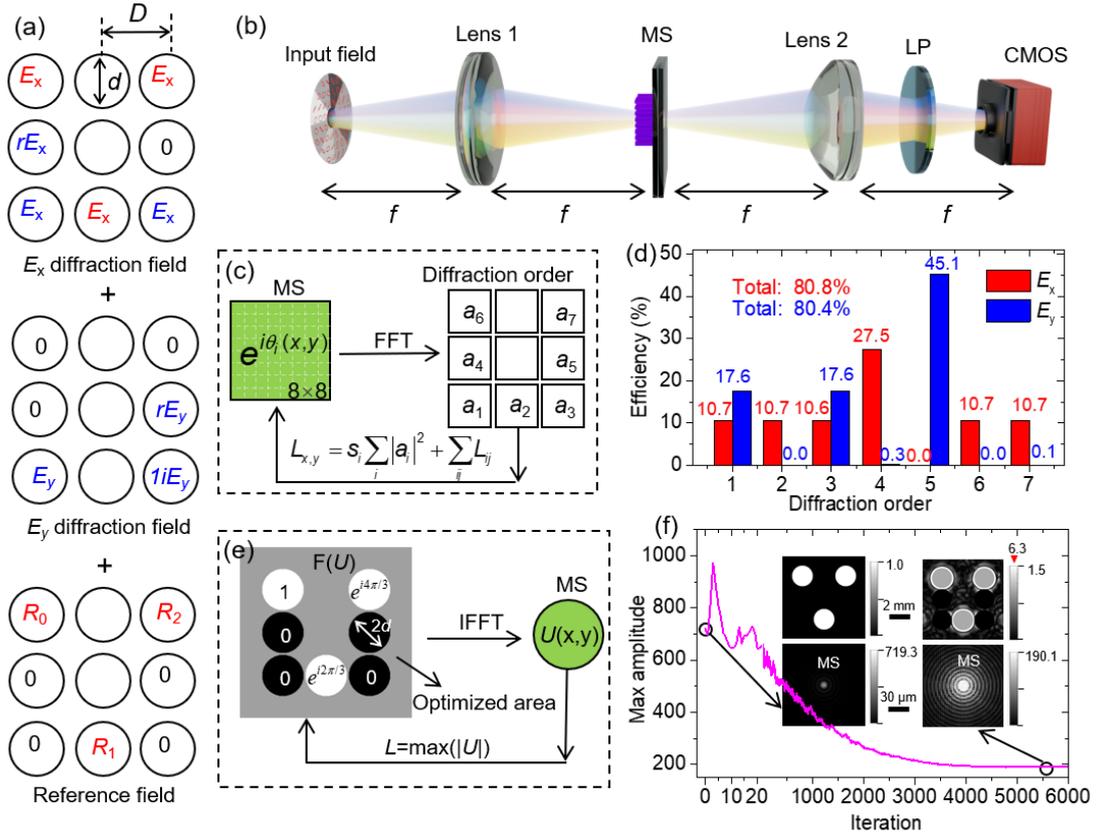

**Figure 2**. Metasurface design for three-parameter optical imaging. (a) Categorization of captured image fields into $E_x$, $E_y$ diffraction field and the reference field $R$. These fields span across seven distinct regions, with certain regions exhibiting null intensities. Each diffraction order image is confined within a circle of diameter $d$, with a center-to-center spacing $D$ between sub-images. The $E_x$ and $E_y$ diffraction fields, located in the mid-left and mid-right regions, are augmented by a factor $r=1.6$. (b) Schematic of the key optical components: two lenses, a metasurface and a linear polarizer. (c) Depiction of the gradient-based optimization aimed to enhance the efficiency of $E_x$ and $E_y$ fields

to their respective desired diffraction orders. The metasurface features an 8×8 periodic pattern with pure phase, where $a_i$ represents the seven orders of interest closest to the zero order. The period of the metasurface $P$ and the center-to-center distance between sub-images $D$ are related by the equation $D=f\lambda/8P$. (d) Optimized efficiency of the seven orders for both $E_x$ field (red bars) and $E_y$ field (blue bars). (e) Illustration of optimizing the intensity of the reference field using a gradient-based approach. Gray areas outside the seven circles indicate the optimization regions. The loss function is established as max($|U|$), with $U$ denoting the metasurface distribution. (f) A plot of loss value (max amplitude of metasurface) against iteration number. Insets display the initial and final optimized distributions of F($U$) (top) and corresponding metasurface distributions (bottom).

The output fields can be categorized into three distinct groups: $E_x$ diffraction field, $E_y$ diffraction field, and the reference field $R_i$, as shown in Figure 2a. These fields within each category are distributed across discrete regions, with certain regions exhibiting null intensities. The maintenance of zero intensity is of importance to prevent any unwanted cross-interference or noise introduction into other categories. Additionally, precise calibration of the amplitudes and phase across different regions in each category is essential. For instance, the $E_x$ fields in the three regions designed to interfere with reference fields must maintain the same complex-amplitude coefficients to extract the phase correctly. Similarly, the two $E_y$ fields designated for interference with $E_x$ fields should exhibit the same amplitude but a phase difference $\pi/2$. The $E_x$ and $E_y$ fields, positioned in the mid-left and mid-right regions, do not interfere with others, which may lead to a lower peak intensity compared to other regions. To ensure consistency of their peak intensities with the rest, the amplitudes of the two fields are augmented by a factor $r$=1.6. Detailed considerations regarding the coefficients of each field are provided Supplementary Section 2.

To realize such output field, we design a MS-based optical system, including two lenses (Lens 1 and Lens 2 with a focal length $f$), a metasurface and a linear polarizer, as illustrated in Figure 2b. The two lenses establish a 4$f$ imaging system, with the metasurface placed at the central Fourier plane, serving to diffract the incoming optical fields into distinct orders. These orders are subsequently captured by the CMOS sensor via Lens 2, resulting in an image segmented into seven distinct sub-images. The linear polarizer is set at 45° relative to the x axis, enabling the interference of the fields between $x$ and $y$ components. Note that these seven diffraction orders have different polarization states and cannot be filtered by the linear polarizer simultaneously.

In the absence of the linear polarizer, the output field generated by the MS-based optical system can be deduced as follows:

$$E_i^{out}(x,y) = F[U_i(\eta,v)] \otimes E_i^{in}(-x,-y) \qquad (3)$$

Here, $F$ represents the Fourier transform operator, and $\otimes$ denotes the convolution operator. The spatial coordinates at the input and output planes are represented by ($x$, $y$), while ($\eta$, $v$) are the spatial coordinates associated with the $x$ and $y$ directions at the metasurface plane. $E_i^{in}$, $E_i^{out}$ and $U_i$ stand for the input field, output field, and the complex amplitude of the metasurface, respectively, with $i$ indicating either the $x$ or $y$ component.

**Metasurface design of $E_x$ and $E_y$ diffraction fields**

To generate the diffracted orders of $E_x$ and $E_y$ fields, the metasurface can be architected with a periodic pattern, incorporating specific Fourier coefficients. A direct formulation of the metasurface can be expressed as the summation of seven plane waves, represented as $\sum_i a_i \exp(ik_x^i \eta + ik_y^i \nu)$. Here, $a_i$ signifies the complex coefficients of different diffraction orders, with certain coefficients equating to zero for null diffractions (Figure 2a). The wavevectors along the $x$ and $y$ directions, $k_x^i$ and $k_y^i$, are determined by the central positions of the seven sub-images ($x_i$, $y_i$) and can be expressed as $k_x^i = k_0 x_i / f$ and $k_y^i = k_0 y_i / f$, with $k_0$ being $2\pi/\lambda$. However, a significant limitation of this approach is the diminishing efficiency as the number of plane waves increases. For instance, when considering $E_x$ and $E_y$ fields with six and three diffraction orders respectively, the overall efficiencies are only 17.4% and 35.2% (Supplementary Section 2).

To address this issue, we employ a gradient-based optimization method to design the metasurface (Figure 2c). The metasurface is designed with 8×8 periodic pattern with pure phase $e^{i\theta_i}$, which is subjected to optimization. The Fourier transformation of the metasurface has 8×8 diffraction orders, and only the seven closest to the zero order are of interest, denoted as $a_i$ (Figure 2c). In the optimization algorithm, the loss functions $L_x$ and $L_y$ are defined as the summation of the absolute square of $a_i$ with prefixed with a ratio $s_i$. The ratio $s_i$ equals to -1 for those associated with existing fields and +1 for those with null intensities. Additionally, an additional loss term $L_{ij}$ is introduced to ensure the consistency of Fourier coefficients across various orders. For example, in loss $L_x$, the Fourier coefficients of $a_2$, $a_6$ and $a_7$ are identical. To enforce this, a term of $|1-a_2/a_6|^2 +|1-a_2/a_7|^2$ is incorporated into $L_{ij}$. More details of the constraints for other Fourier coefficients and the optimization method are provided in Supplementary Section 2. Following this optimized approach, the efficiencies for $E_x$ and $E_y$ fields are significantly enhanced to 80.8% and 80.4% respectively, as illustrated in Figs. 2d.

**Metasurface design of reference field**

To generate the reference field, we employ the existing optical field to establish the reference one. In contrast to conventional interferometric phase imaging, which necessitates an additional optical beam from a known laser source as the reference field, our approach presents distinct advantages. It not only simplifies the optical measurement but also proves effective in scenarios where obtaining a reference beam from an unknown source is impractical. The crucial concern here is how to consistently maintain uniform references and null fields within specified regions of interest (bottom panel of Figure 2a), regardless of variations in the input optical field. Referring to Equation 3, we configure the Fourier transform of the metasurface ($F(U)$) to exhibit patterns similar to those of reference fields, but with the diameter of the seven circles augmented to $2d$, depicted in Figure 2e. This design, when convoluted with an arbitrary field, generates the desired field distributions within the designated regions (see Supplementary Section 3 for details).

The complex amplitude of the metasurface $U$, can be determined by performing an inverse Fourier transformation on the $F(U)$ pattern. With a unity magnitude in the three circular regions and zeros elsewhere in the Fourier plane, the maximal amplitude of the metasurface $U$ is computed to be 719.3 (inset in Figure 2f). The distribution of the metasurface with high amplitude values predominantly centers around a narrow central region. However, such straightforward operation can lead to low reference intensity when the metasurface amplitude is normalized, thereby compromising interference contrast. It is note that only the seven circular regions with diameter $d$ contribute to field retrieval. Therefore, in the $F(U)$ pattern, areas outside these seven circles with diameter $2d$ (gray regions in Figure 2e) can be optimized to minimize the maximal amplitude of the metasurface, thereby increasing the intensity of reference field.

We employ a gradient-based algorithm (see Supplementary Section 3 for details), using a loss function of $\max(|U|)$, as depicted in Figure 2e. The initial distribution of $F(U)$ begins with a unity magnitude in the three circular regions and zeroes elsewhere, consistent with the approach previously used. Figure 2f shows the loss value as a function of the iteration which reaches convergence after 6000 iterations. Post-convergence, the peak amplitude of the metasurface is reduced to 190.1 from an initial 719.3, signifying a 14.3-fold enhancement in reference intensity. In this scenario, the fields in the optimized area of $F(U)$ pattern manifest a varied distribution, and the high-amplitude values of metasurface pattern are more widely dispersed compared to the initial design (inset of Figure 2f). We verify the generation of reference field and find that the intensities and phases can be uniformly manifested within the delineated three circular regions (see Supplementary Section 3). In this case, the interference between the diffracted $E_x$ field and reference field yields a max/min intensity ratio of 5.14, a sufficient contrast for phase extraction (see Supplementary Section 3).

**Metasurface design of Jones matrix**

We proceed to the design of Jones matrix metasurface, aiming to realize all the three functions described above. For *xx* component of the Jones matrix, we set $J_{xx} = 0.5e^{i\theta_x} + 0.5U_{norm}$, where $\theta_x$ represents the optimized metasurface phase for $E_x$ diffraction field and $U_{norm}$ is the normalized metasurface pattern for reference fields. To ensure coefficient consistency between $E_x$ and $E_y$ fields, the *yy* component is set as $J_{yy} = 0.39e^{i\theta_y}$, with $\theta_y$ denoting the optimized metasurface phase for $E_y$ diffraction field. The factor 0.39 is deduced from $0.5/\sqrt{1.645}$, where 1.645 (17.6%/10.7%, Figure 2d) is the efficiency ratio of the single diffraction order between $E_x$ and $E_y$ fields. Additionally, a reference field term of $0.61U_{norm}$ is incorporated into $J_{yy}$ to ensure unity peak amplitude. In this design, reference fields are encoded in both *x* and *y* polarizations, and can both interfere with $E_x$ fields, supported by the linear polarizer. For certain cases, the induced intensity of reference field may be insufficient to achieve a suitable contrast for interference. This challenge can be effectively addressed through several strategies, including adjusting the linear polarizer axis, modifying the field of view size, or rotating the metasurface (see Supplementary Section 3).

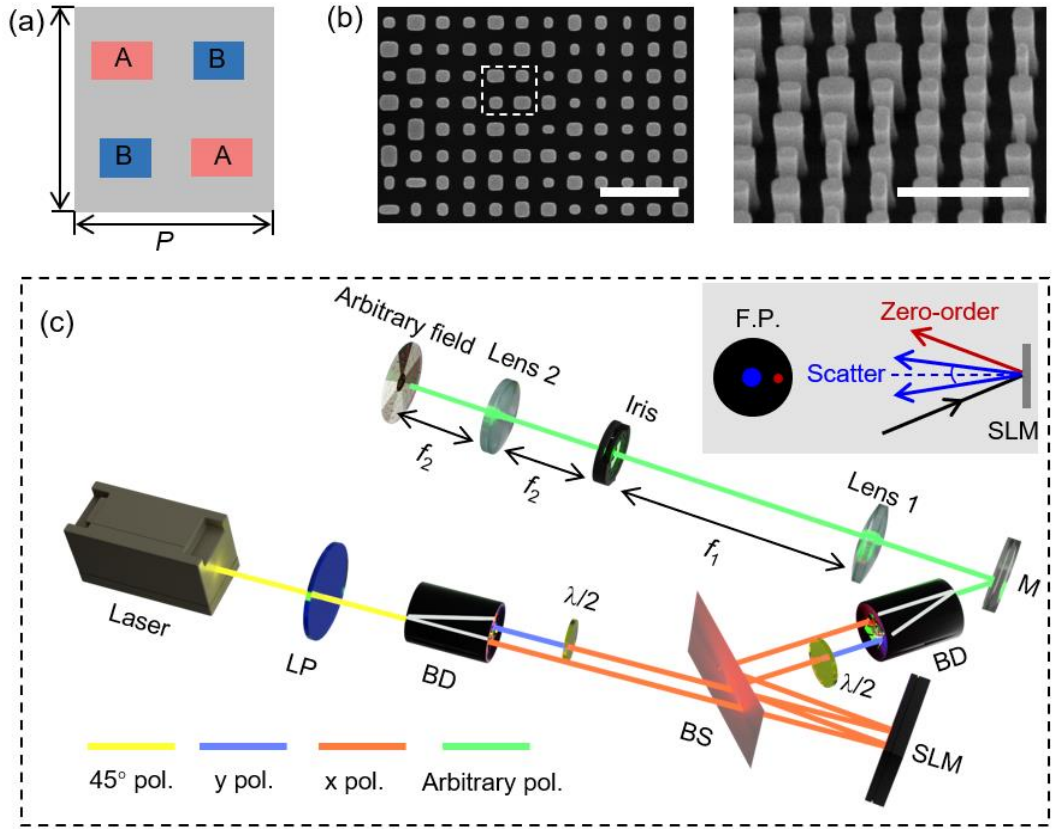

**Figure 3**. Design of Jones matrix metasurface and optical setup to generate arbitrary field. (a) Diagram illustrating the tetratomic micropixel unit of the metasurface designed for Jones matrix construction. Schematic of tetratomic micropixel unit of metasurface for constructing the Jones matrix. Each pixel contains of two unique rectangular nanoblocks labeled as A and B. (b) Scanning electron microscopy (SEM) images of the fabricated metasurface (partial view). The white rectangle indicates a micropixel with four nanoblocks. Scale bar: 1 μm. (c) Optical setup to generate optical field with arbitrary intensity, phase, and polarization distribution. The inset details the positions in Fourier plane (F.P.) of the incident zero order and the field scattered by SLM. (LP linear polarizer, BD, beam displacer, $\lambda/2$: half waveplate, $\lambda/4$: quarter waveplate, BS: beam splitter, M: mirror)

We employ a tetratomic micropixel consisting of two distinct rectangle nanoblocks, termed meta-atoms *A* and *B*, for constructing the Jones matrix[36, 37] (Figure 3a). Given that the off-diagonal entries of the Jones matrix are zero, both meta-atoms possess zero rotational angles. The phase shift values for the two meta-atoms along the *x* and *y* axes can be directly deduced from the established Jones matrix. Full wave finite-difference time-domain (FDTD) simulations are performed to create a library detailing the transmission and phase shifts dependent on the nanoblock's transverse dimensions. The desired phase shifts are obtained by appropriately choosing the transverse dimensions of the nanoblocks. The micropixel's period is $P$=700 nm and is designed for $\lambda$=780 nm. More details are provided in Supplementary Section 4.

The metasurface sample, designed with a diameter of 600 μm, is fabricated on 600 nm-height crystal silicon layer that is transferred on glass substrate. The patterns are then defined by electron beam lithography and reactive ion etching process. The SEM images, showing both the top and oblique perspectives, are illustrated in Figure 3b. The details of the fabrication procedure can be found in "Materials and Methods" section.

**Generation of arbitrary optical field and experimental verification**

Generally, any optical field distribution, characterized by arbitrary intensity, phase, and polarization, can be decomposed into two independent optical fields polarized along *x* and *y* directions, each possessing distinct amplitude and phase distributions. Therefore, we can construct such optical field by modulating its two polarized components separately. The experimental setup is shown in Figure 3c. In this setup, the incident light is initially split into horizontal and vertical polarization components, which then illuminate the left and right halves of the spatial light modulator (SLM) screen. Here, the pure phase pattern in SLM can be used to encode arbitrary independent amplitude and phase information [38] (see Supplementary Section 5 for details). Subsequently, these two beams are coherently superimposed, generating an arbitrary optical field with predefined intensity, phase, and polarization distributions. By illuminating the SLM at a slight oblique angle, the unwanted zero order can be shifted away from the Fourier plane's center (red point in the inset of Figure 3c), which can be easily filtered by the iris. The details of the setup can be found in "Materials and Methods" section.

The described optical setup provides the flexibility to generate arbitrary optical field with well-defined distribution, which serves as a benchmark to assess the accuracy of our proposed MS-based imaging system. To illustrate this capability, we have designed three general optical fields, as presented in the first column of Figure 4. These fields exhibit varying intensity and phase distributions in $x$- and $y$ components (i.e., $|E_x|^2$, $|E_y|^2$, $\varphi_x$ and $\varphi_x$-$\varphi_y$), including special phase distributions like vortex patterns that are unfeasible with TIE method. The results of our measurement, illustrated in the second column of Figure 4, reveal seven distinct sub-images. The zero-order image in the center has been manually obscured to avoid interference with observation.

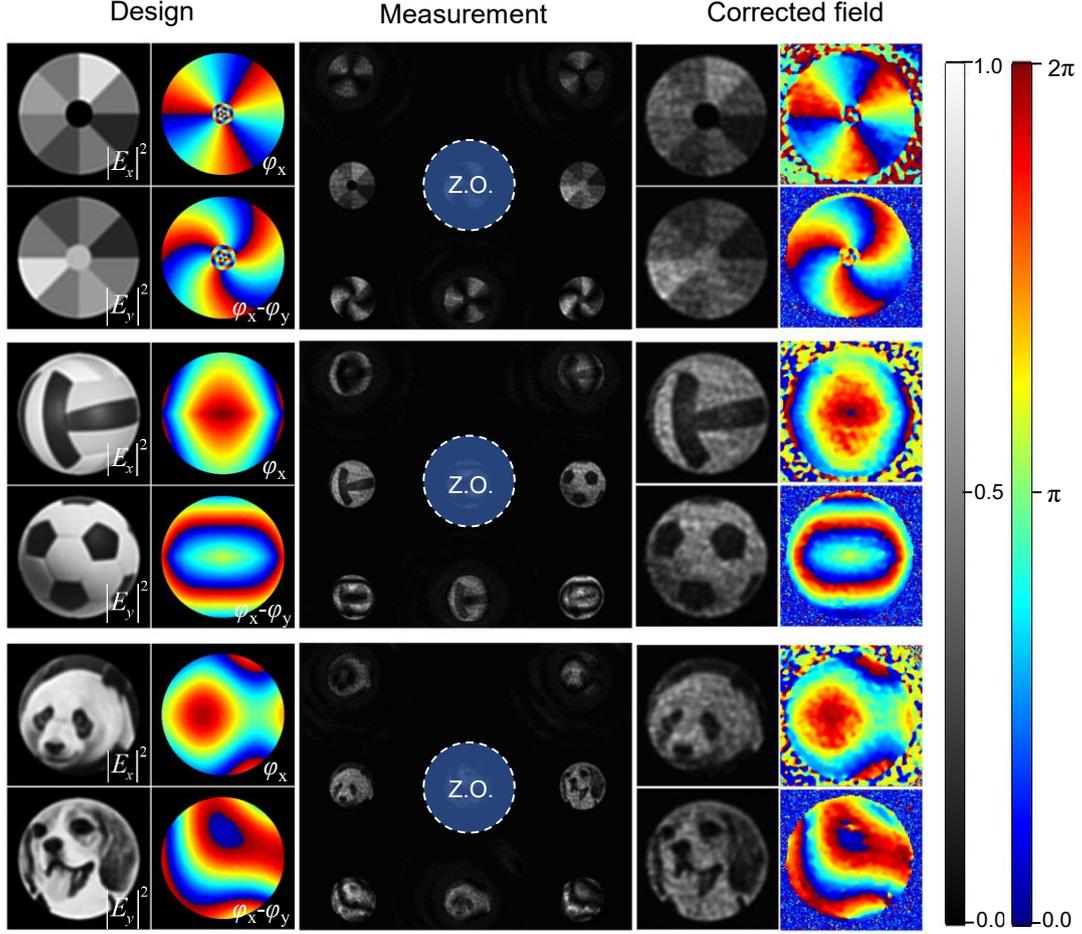

**Figure 4**. Experimental verification of single-shot three-parameter optical imaging. The leftmost column displays three distinct optical fields, each characterized by unique intensity and phase distributions for the *x*- and *y* components (i.e., $|E_x|^2$, $|E_y|^2$, $\varphi_x$ and $\varphi_x$-$\varphi_y$). The center column reveals the measurement results. The zero-order (Z.O.) images are intentionally concealed to enhance visibility. The right column presents the fields retrieved post image correction.

When the incident field is diffracted by the metasurface into several orders, it creates certain angles with the lens's principal axis, leading to imaging aberrations. This is especially apparent in the distortion seen in the four corner sub-images in measurement. To rectify this distortion, an image correction process is employed for all seven sub-images, with specifics outlined in Supplementary Section 6. The corrected field distributions for all three cases are displayed in the third column of Figure 4, all agreeing well with our original designs. The observed speckles in the measured images primarily arise from the coherence of the laser beam and possibly the inherent limitations of encoding complex-amplitude using SLM, and thus are unrelated to the metasurface design (Supplementary Section 5). The efficiency of metasurface for imaging is addressed with the details in Supplementary Section 7. Since the incident field on the metasurface represents the Fourier transformation of the original incident pattern, the efficiency varies and depends on the original field distribution. The measured efficiency for the three types of optical fields ranges from 17.5% to 22.4%, a value nearly acceptable for practical use.

**CONCLUSIONS**

In summary, we have demonstrated a MS-based single-shot imaging system capable of capturing optical fields with arbitrary intensity, phase, and polarization distributions. The primary optical setup employs a 4$f$ system with the metasurface placed at the Fourier plane. The system features two lenses with identical focal lengths of 15 mm and a metasurface diameter of 600 μm, resulting in a numerical aperture (N.A.) of 0.02 and a spatial resolution of approximately 19 μm. This spatial resolution can be adjusted by changing the metasurface diameter or the focal length of Lens 1. Additionally, the simultaneous capture of seven diffraction images reduces the field of view (FOV) to about 1 mm. The FOV can also be modified by altering the focal length of Lens 1. There is always a trade-off between spatial resolution and FOV: increasing the focal length of Lens 1 enlarges the FOV but reduces spatial resolution, and vice versa. The optimal balance between these parameters depends on the specific imaging application requirements.

For the metasurface design, the Jones matrix is carefully designed to generate the desired reference field and diffract the incident *x* and *y*-polarized fields into distinct orders with specific complex-amplitude coefficients. Notably, we have incorporated optimizations into the metasurface design to maximize the efficiency of both the reference field generation and the diffraction of *x* and *y*-polarized light. The design in our work enables the imaging of arbitrary fields with single shot, effectively addressing the limitations inherent in traditional methods. Specifically, it eliminates the need for multiple measurements of the three optical parameters, reduces reliance on a laser source for reference beam in conventional phase measurements, and overcomes the incapacity of the traditional DPI method to measure phase distributions of arbitrary fields, et. al. We believe that our work represents a significant advancement in optical field imaging, offering a highly efficient and versatile solution with broad applicability including microscopy, materials science and optical communications.

## MATERIALS AND METHODS

**Metasurface fabrication.** A commercial silicon-on-insulator (SOI) wafer, initially having a device layer thickness of 1200 nm, was initially transferred to a glass substrate through adhesive wafer bonding and deep reactive ion etching (DRIE) techniques[39]. Subsequently, the thickness of the device layer was further reduced to 600 nm by utilizing inductively coupled plasma (ICP) etching. Following this, a 300 nm-thick layer of hydrogen silsesquioxane (HSQ) was spin-coated onto the substrate at a rate of 4000 revolutions per minute and then baked on a hot plate at 90 °C for 5 minutes. Next, a 30 nm-thick layer of aluminum was deposited onto the HSQ layer through thermal evaporation to serve as the charge dissipation layer. The metasurface mask on the HSQ layer was created using electron beam lithography at an acceleration voltage of 30 kV. After the exposure, the aluminum layer was removed using a 5% phosphoric acid solution, and the resist was developed with tetramethylammonium hydroxide (TMAH) for 2 minutes at room temperature (25 °C). Subsequently, inductively coupled plasma-reactive ion etching (ICP-RIE) was employed to transfer the pattern into the silicon film. Finally, the samples were immersed in a 10% hydrofluoric (HF) acid solution for 15s to eliminate any residual HSQ mask. The samples were then cleaned with deionized water and dried using nitrogen gas.

**Generation of arbitrary field distributions.** The experimental setup for arbitrary field generation is shown in Figure 3c. The linear polarization of an input laser beam is oriented at 45° after passing through a linear polarizer. A beam displacer (BD40, Thorlabs) subsequently splits this beam into horizontal and vertical polarization components. The two components, propagating parallelly, illuminate the left and right halves of the SLM (PLUTO-2-NIR-011, Holoeye) screen, where distinct phase patterns are imposed to encode both amplitude and phase information[38]. Since the SLM exclusively responds to $x$-polarization, a half-wave plate is positioned before the SLM to convert the incident $y$-polarization to $x$-polarization. Subsequently, one of the reflected $x$-polarized beams is transformed to $y$-polarization via another half-wave plate. It is superposed coaxially with the remaining $x$-polarized field using a second beam displacer, resulting in a versatile optical field with specified intensity, phase, and polarization distributions. The generated field is then directed through a lens pair, forming a 4$f$ optical system, with an iris placed at the Fourier plane to filter the zero order of the incident laser. We purposely set the laser to strike the SLM at a slight oblique angle. As a result, the zero order shifts away from the Fourier plane's center (highlighted by the red point in the inset of Figure 3c), enabling easy filtering by the iris. This oblique incidence imposes a gradient phase on the SLM, which can be compensated by the applied phase of the SLM to align the center of the angular spectrum of the generated fields with the center in the Fourier plane (marked by the blue point in the inset of Figure 3c). The focal lengths of the two lenses ($f_1$ =150 mm and $f_2$=50 mm) are selected to ensure that the resultant optical field adopts a circular shape with a diameter smaller than $d$=1 mm.

## AUTHOR CONTRIBUTIONS

Y.B. conceived the idea, conducted the numerical simulations, performed the experiments, and wrote the manuscript. Y.B. and B.L supervised the project.

Yanjun Bao*, and Baojun Li*

Guangdong Provincial Key Laboratory of Nanophotonic Manipulation, Institute of Nanophotonics, College of Physics & Optoelectronic Engineering, Jinan University, Guangzhou 511443, China

*Corresponding Authors: Y. Bao (yanjunbao@jnu.edu.cn), B. Li (baojunli@jnu.edu.cn)

**Table of Contents**

**Section 1. Parameter retrieval with seven sub-images**

In this section, we present the derivation of intensity, phase, and polarization distributions based on the captured seven sub-images. We assume that the incident fields are represented as follows: for x-polarization, $E_x(x, y) = A_x(x, y)e^{i\varphi_x(x,y)}$, and for y-polarization, $E_y(x, y) = A_y(x, y)e^{i\varphi_y(x,y)}$. Thus, the measurement of intensity, phase, and polarization is equivalent to the measurement of full field distributions of $A_x$, $A_y$, $\varphi_x$, and $\varphi_y$.

Among the seven sub-images, three correspond to the interference patterns between $E_x(x, y)$ and reference fields $R_m = A_r e^{i2m\pi/3}$, where $A_r$ represents a uniform amplitude, and $m$ takes on values 0, 1, and 2. These interference patterns yield intensities that can be expressed as:

$$I_m(x, y) = \left| E_x(x, y) + A_r e^{i2m\pi/3} \right|^2$$
$$= A_x^2(x, y) + A_r^2 + 2 A_x A_r(x, y) \cos\left[\varphi_x(x, y) - 2m\pi/3\right] \quad \text{(S1)}$$

Consequently, we have

$$I_0 - I_1 = 2 A_x A_r \left[\cos\varphi_x - \cos(\varphi_x - 2\pi/3)\right]$$
$$= -2\sqrt{3} A_x A_r \left(\frac{1}{2}\sin\varphi_x - \frac{\sqrt{3}}{2}\cos\varphi_x\right) \quad \text{(S2)}$$

$$I_0 - I_2 = 2 A_x A_r \left[\cos\varphi_x - \cos(\varphi_x - 4\pi/3)\right]$$
$$= -2\sqrt{3} A_x A_r \left(-\frac{1}{2}\sin\varphi_x - \frac{\sqrt{3}}{2}\cos\varphi_x\right) \quad \text{(S3)}$$

Then we can obtain:

$$\cos\varphi_x = \frac{1}{6 A_x A_r}(2I_0 - I_1 - I_2) \quad \text{(S4)}$$

$$\sin \varphi_x = \frac{1}{2\sqrt{3}A_x A_r}(I_1 - I_2) \tag{S5}$$

The phase of $E_x$ can be directly obtained as:

$$\varphi_x = \mathrm{atan2}\left(\sqrt{3}(I_1 - I_2), 2I_0 - I_1 - I_2\right) \tag{S6}$$

This equation corresponds to Equation 1 of the main text.

The other four sub-images are the intensities of $E_x$, $E_y$, $E_x+1iE_y$, and $E_x+E_y$, labeled as $I'_0$, $I'_1$, $I'_2$ and $I'_3$, respectively. Therefore, we have

$$I'_0 = |E_x|^2 = A_x^2$$

$$I'_1 = |E_y|^2 = A_y^2$$

$$I'_2 = |E_x + 1iE_y|^2 = A_x^2 + A_y^2 + 2A_x A_y \sin(\varphi_x - \varphi_y)$$

$$I'_3 = |E_x + E_y|^2 = A_x^2 + A_y^2 + 2A_x A_y \cos(\varphi_x - \varphi_y)$$

Based on the four equations, we can obtain:

$$\varphi_x - \varphi_y = \mathrm{atan2}\left(I'_2 - I'_0 - I'_1, I'_3 - I'_0 - I'_1\right) \tag{S7}$$

This corresponds to Equation 2 of the main text. In the actual design, the amplitudes of the two fields $E_x$ and $E_y$ are increased by a factor $r=1.6$ to ensure consistency of their peak intensities with the rest. Therefore, the phase difference becomes:

$$\varphi_x - \varphi_y = \mathrm{atan2}\left(I'_2 - I'_0/r^2 - I'_1/r^2, I'_3 - I'_0/r^2 - I'_1/r^2\right) \tag{S8}$$

**Section 2. Multi-order diffraction design with gradient-based optimization**

In this section, our aim is to design a metasurface capable of producing seven diffraction sub-images of the $E_x$ and $E_y$ fields with maximal efficiency. When the input field $E_i^{in}(x, y)$ passes through Lens 1, the field at the metasurface plane is represented as:

$$E_i^m(\eta,v) = \frac{\exp(i2kf)}{i\lambda f} \int_{-\infty}^{+\infty}\int_{-\infty}^{+\infty} E_i^{in}(x,y)\exp\left[-i2\pi\left(x\frac{\eta}{f\lambda}+y\frac{v}{f\lambda}\right)\right]dxdy$$

$$\propto F\left[E_i^{in}(x,y)\right] \tag{S9}$$

Here, $f$ denotes the focal length of both Lenses 1 and 2, $F$ represents the Fourier transformation. Assuming the complex-amplitude of the metasurface pattern to be $U_i$, the light transmitted through Lens 2 and imaged at the focal distance can be defined as:

$$E_i^{out}(x,y) = \frac{\exp(i2kf)}{i\lambda f} \int_{-\infty}^{+\infty}\int_{-\infty}^{+\infty} E_i^m(\eta,v)U_i(\eta,v)\exp\left[-i2\pi\left(\eta\frac{x}{f\lambda}+v\frac{y}{f\lambda}\right)\right]d\eta dv$$

$$\propto F\left[E_i^m(\eta,v)U_i(\eta,v)\right]$$

$$\propto F\left\{F\left[E_i^{in}(x,y)\right]U_i(\eta,v)\right\}$$

$$= F\left[U_i(\eta,v)\right] \otimes E_i^{in}(-x,-y) \tag{S10}$$

This corresponds to Equation 3 of the main text. To produce seven sub-images with a coefficient of $a_i$, a straightforward method is:

$$U_i(\eta,v) = \sum_i a_i \exp(ik_x^i\eta + ik_y^i v) \tag{S11}$$

termed as the direct plane-wave summation method. This gives:

$$F\left[U_i(\eta,v)\right] \otimes E_i^{in}(-x,-y)$$

$$= \left[\sum_i a_i \delta(f\lambda k_x^i/2\pi, f\lambda k_y^i/2\pi)\right] \otimes E_i^{in}(-x,-y)$$

$$= a_i \sum_i E_i^{in}(-x+f\lambda k_x^i/2\pi, -y+f\lambda k_y^i/2\pi) \tag{S12}$$

This represents a summation of multiple $E_i^{in}$ at distinct regions, each with a coefficient $a_i$. Giving that the center of the sub-images is represented by ($x_i$, $y_i$), we can determine:

$$k_x^i = k_0 x_i / f$$

$$k_y^i = k_0 y_i / f$$

where $k_0=2\pi/\lambda$. However, an inherent limitation of this approach is the decreased efficiency with an increasing number of plane waves. Owing to the absence of gain in the metasurface, the maximal amplitude cannot surpass 1.0. After normalizing the metasurface pattern $U_i(\eta,\nu)$, the efficiency of the metasurface can be expressed as:

$$T = \frac{\sum_i |a_i|^2}{\left(\sum_i |a_i|\right)^2} \tag{S13}$$

For our design, we have $a_1=a_2=a_3=a_6=a_7=1.0$, $a_4=1.6$, $a_5=0.0$ for $E_x$ field and $a_1=1.0$, $a_2=a_4=a_6=a_7=0.0$, $a_3=1i$, $a_5=1.6$ for $E_y$ field. The obtained efficiencies are $T_x=17.36\%$ and $T_y=35.19\%$.

To address the aforementioned efficiency concerns, we employed gradient-based optimization for designing the metasurface pattern. If we assume the unit period of the metasurface to be $P$ and the metasurface pattern exhibits a pure phase of $e^{i\theta_i}$ ($i=x, y$) with a periodicity of $L=NP$ (where $N=8$) along both $x$ and $y$ axes, then:

$$F[U_i(\eta,\nu)] = \sum_{mn} a_{mn}\delta(x-\frac{mf\lambda}{8P}, y-\frac{nf\lambda}{8P}) \tag{S14}$$

where $a_{mn} = \frac{1}{L^2}\int_0^L\int_0^L U_i(\eta,\nu)\exp\left[-i2\pi\left(\eta\frac{x}{f\lambda}+\nu\frac{y}{f\lambda}\right)\right]d\eta d\nu = \frac{1}{N^2}\text{FFT}(U)$ with FFT symbolizing the fast Fourier transformation.

Subsequently, $E_i^{out}(x, y)$ can be expressed as:

$$E_i^{out}(x,y) = \sum_{mn} a_{mn}\delta(x-\frac{mf\lambda}{8P}, y-\frac{nf\lambda}{8P}) \otimes E_i^{in}(-x,-y)$$

$$= a_{mn}\sum_{mn} E_i^{in}(-x-\frac{mf\lambda}{8P}, -y-\frac{nf\lambda}{8P}) \tag{S15}$$

Here, $a_{mn}$ represents for the coefficients of the sub-images. The sub-images of interest are the nearest seven to the zero order, leading to the center-to-center distance between sub-images $D$ as $D=f\lambda/8P$. In the subsequent discussions, the coefficients of these seven orders $a_{mn}$ are denoted as $a_i$ for brevity.

Our approach aims to enhance the overall efficiency of the seven orders while maintaining consistent coefficient ratios among them. To achieve this goal, our loss function comprises two essential components: efficiency and constraint. As the algorithm is to minimize the loss function, the efficiency term for the $E_x$ and $E_y$ fields is expressed as:

$$L_x = -|a_1|^2 - |a_2|^2 - |a_3|^2 - |a_4|^2 - |a_6|^2 - |a_7|^2 + |a_5|^2 = -\sum_{i\neq 5}|a_i|^2 + |a_5|^2 \quad (S16)$$

$$L_y = -|a_1|^2 - |a_3|^2 - |a_5|^2 + |a_2|^2 + |a_4|^2 + |a_6|^2 + |a_7|^2 = -\sum_{i=1,3,5}|a_i|^2 + \sum_{i=2,4,6,7}|a_i|^2 \quad (S17)$$

where negative signs are associated with existing fields, while positive signs correspond to null intensities.

Furthermore, it is essential to maintain specific ratios between different coefficients, which can be summarized as follows: $a_1=a_3$, $a_2=a_6=a_7$, $|a_1|=|a_2|$, $|a_4|=r|a_2|$ for $E_x$ field, and $a_3=1ia_1$, $|a_5|=r|a_1|$ for $E_y$ field.

Consequently, the loss due to ratio constraints can be expressed as:

$$\sum_{ij} L_{ij} = |1-a_2/a_6|^2 + |1-a_2/a_7|^2 + |1-a_3/a_1|^2 + (1-|a_1/a_2|)^2 + (r-|a_4/a_2|)^2 \quad (S18)$$

$$\sum_{ij} L_{ij} = |1i - a_3/a_1|^2 + (r-|a_5/a_1|)^2 \quad (S19)$$

which apply to the $E_x$ and $E_y$ fields, respectively. The total loss function combines the efficiency term and the ratio constraints term.

We use PyTorch 1.13 for gradient calculation, which automatically computes gradients. Our training starts with a randomly initialized phase and uses the "Adam" optimizer with an initial learning rate of 3e-2. we fine-tune this learning rate dynamically using 'ReduceLROnPlateau' with a patience of 30, a threshold of 0.01, and a decay rate of 0.9 to ensure efficient model convergence and performance.

Supplementary figure 1 depicts the loss as a function of the iteration. Notably, all constraint terms and the efficiency associated with null fields converge to zero, confirming the preservation of the desired coefficient ratios. The final efficiencies of the $E_x$ and $E_y$ fields significantly improve to 80.8% and 80.4%, respectively. A comparison of the efficiency for various orders with direct plane-wave summation and optimization is shown in Figs. S1c-d.

Furthermore, for polarization retrieval, it is necessary that the $a_1$ coefficients of $E_x$ field $a_{1x}$ and $E_y$ field $a_{1y}$ should be identical. To achieve this, a coefficient $c=a_{1x}/a_{1y}$ is multiplied by the metasurface pattern of $U_y$. In our main text, when we set $U_x = 0.5e^{i\theta_x}$, we have $U_y = 0.5ce^{i\theta_y} = 0.39e^{i\theta_y}$, where the phase of $c$ is incorporated to $e^{i\theta_y}$.

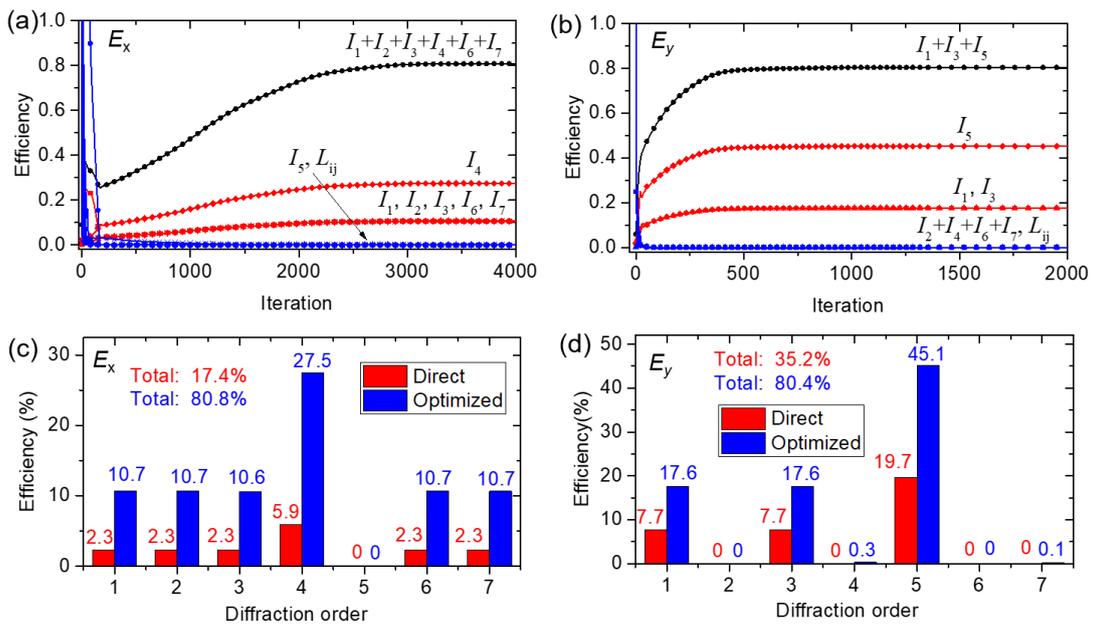

Supplementary Figure 1. Optimization of multi-diffraction orders with constrained Fourier coefficients. (a) Loss value plotted against the number of iterations for $E_x$ field diffraction. (b) Loss value plotted against the number of iterations for $E_y$ field diffraction. The black lines represent the total efficiency, while the red line represents the individual efficiency of the order of interest. The blue lines correspond to the order that is intended to be minimized (zeroed). (c, d) Efficiency comparisons for various orders using direct plane-wave summation (red bars) and optimization (blue bars) for both Ex field (c) and Ey field (d).

**Section 3. Optimization of reference field**

The phase retrieval of the $E_x$ field necessitates the use of three uniform reference fields with equidistant phases in three distinct regions, each having a diameter of $d$ (refer to supplementary fig. 2a). As per Eq. S10, the reference field is derived from the input field. The critical question here is how to generate these desired reference fields consistently, regardless of variations in the input field.

Based on the properties of convolution, it becomes evident that if the Fourier transformation of the metasurface pattern exhibits uniform reference fields with $e^{i2\pi m}$ at three distinct distance regions, each with a diameter of $2d$, then it will generate the desired reference field within the three circles with a diameter of $d$ (supplementary fig.

2b). The generated uniform complex amplitude can be calculated as:

$$\int_{r<d/2} e^{i2\pi m/3} E_i^{in} dxdy = e^{i2\pi m/3} \int_{r<d/2} E_i^{in} dxdy = B_0 e^{i\varphi_0} e^{i2\pi m/3} \tag{S20}$$

Here, we assume that the input field $E_i^{in}$ has a field shape with a diameter of $d$ and is zero outside. Consequently, the bounds of the integral above are confined to the circular areas. Furthermore, assuming zero field values in the other four circular regions ensures that the generated reference fields remain zero, avoiding interference with other fields. The metasurface pattern can be directly obtained by performing an inverse Fourier transformation of the $F(U)$ pattern. Assuming zero fields in the regions outside the three circular regions (each with a diameter of $2d$), the maximum amplitude of the metasurface is calculated to be 719.3.

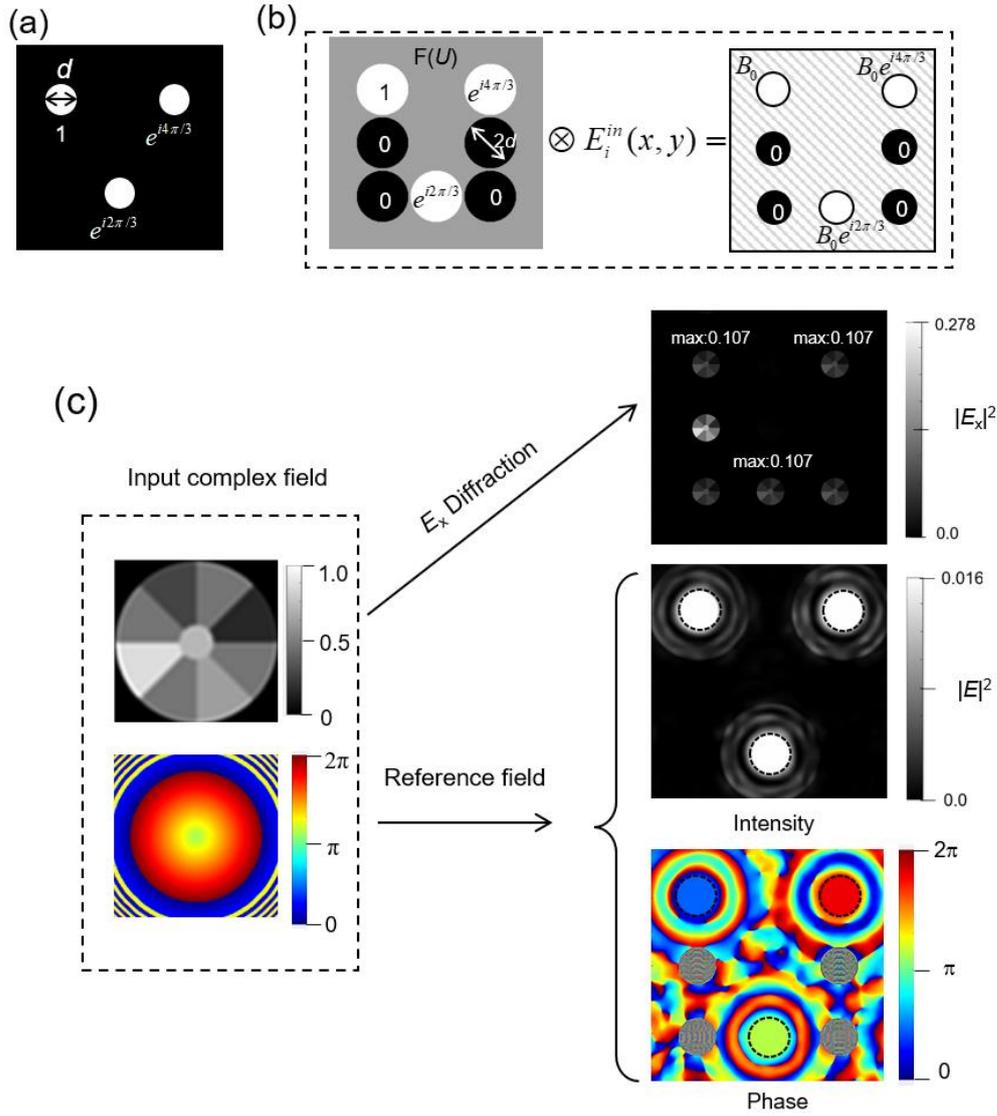

Supplementary Figure 2. Generation of reference field from arbitrary input field. (a) Illustration of an ideal reference field comprising three circular regions, each with a diameter of $d$. (b) The generation of a reference field from an arbitrary input field involves the convolution of the input field with the Fourier transformation of the metasurface pattern $F(U)$. The uniformity of the reference field is ensured by the uniform $F(U)$ pattern within seven circular regions, each having a diameter of $2d$. (c) Generation of the $E_x$ diffraction and the reference field for an input complex field.

To increase the amplitude of the reference field, we must minimize the maximum amplitude of the metasurface pattern. The complex amplitude in the areas outside the seven circular regions is assumed to be optimized (gray areas in supplementary fig. 2b), and the loss function max($|U|$) is employed, where $U$ represents the complex amplitude of the metasurface. The gradient calculations are performed utilizing PyTorch 1.13, with the parameters consistent with those described in the aforementioned diffraction optimization process. The key distinction lies in the choice of the learning rate, as it significantly influences the maximal amplitude of the metasurface, as shown in supplementary fig. 3. Notably, we have observed that a higher learning rate results in a diminished maximal amplitude of the metasurface. Nevertheless, this also leads to an increase in amplitude in the peripheral regions outside the seven circular areas, consequently giving rise to a larger stray background field. To balance, we have selected a learning rate of 0.06. Post-convergence, the peak amplitude of the metasurface is reduced to 190.1.

A complex field (supplementary fig. 2c) is introduced to examine the reference field and the diffraction field $E_x$. For a normalized metasurface designed for $E_x$ field diffraction, it generates seven sub-images with a maximum intensity of 0.278. Among these sub-images, the three interfering with reference field exhibit a maximum intensity of 0.107 (supplementary fig. 2c). In comparison, a normalized metasurface designed for use as a reference field induces a uniform intensity of 0.016 within the three circular regions. Considering all possibilities, the interference between these two fields results in a maximum intensity of $\left|\sqrt{0.107}+\sqrt{0.016}\right|^2 = 0.206$ and a minimal intensity of $\left|\sqrt{0.107}-\sqrt{0.016}\right|^2 = 0.04$, yielding a maximum-to-minimum intensity ratio of 5.14. For the reference field, it can be found that the intensities and phases are uniformly manifested within the delineated three circular regions (denoted by circles of diameter *d*). The phases are in perfect alignment with the design values of $2m\pi/3$. While peripheral stray fields are present around the targeted regions, they do not affect the measurement of the optical fields.

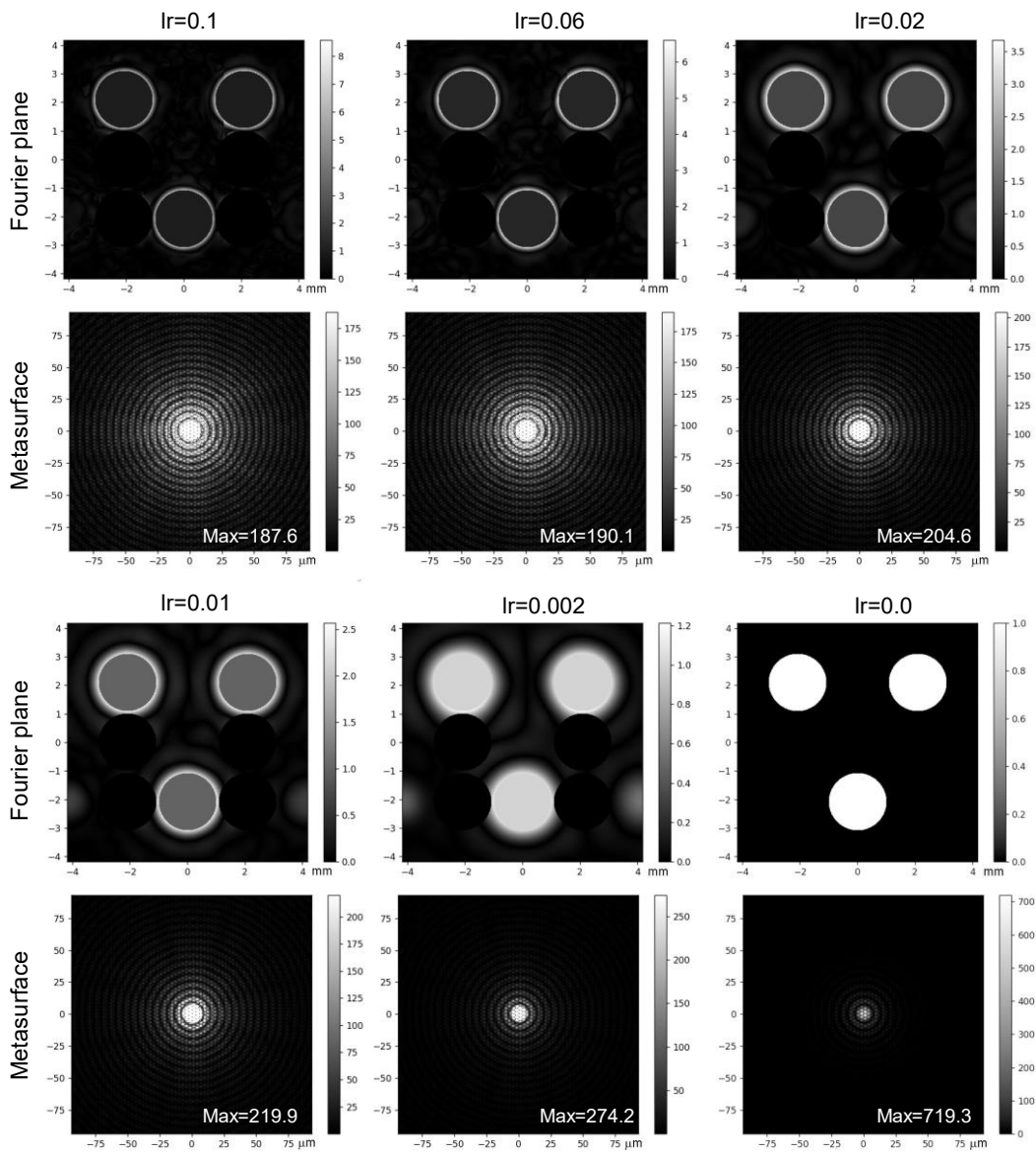

Supplementary Figure 3. The optimized metasurface pattern (bottom) the corresponding Fourier transformation (top) for the generation of reference field with different learning rates. The maximal amplitude of metasurface pattern is indicated in each panel.

It is essential to highlight that the intensity of the reference field is highly dependent on the input field pattern. In certain cases, it may result in significantly lower intensity compared to the $E_x$ field. To address this issue, we propose the following strategies:

1. Adjusting the linear polarizer axis: Assuming that the linear polarizer is initially set at an angle $\theta$ (defaulting to 45°), the complex amplitude of the $E_x$ field for interference $E_x^{in}\cos\theta$. In practical design, the reference field is encoded in both $x$ and $y$ polarizations, and thus, the reference field for interference becomes $B_x e^{i\varphi_x}\cos\theta + B_y e^{i\varphi_y}\sin\theta$, where $B$ and $\varphi$ represent the amplitude and phase of the uniform reference field. It is evident that increasing the angle $\theta$ will decrease the $E_x$ field, resulting in an enhanced contrast of the interference pattern.

2. Modifying the field of view size: In experiments, we have employed an iris to control the size of the input field within a diameter less than $d$=1 mm. Since the complex amplitude of the reference field is obtained by integrating the input field over its area where the existing field is present, the reference field's intensity can be adjusted by modifying the diameter of the iris. Therefore, we can choose an appropriate iris diameter to enhance the reference field intensity.

3. Rotate the metasurface angle. By rotating the metasurface with an angle $\theta$, we effectively perform a coordinate transformation of the input field. In the metasurface's coordinate system, the input $E_x$ and $E_y$ fields become $E_x^{in}\cos\theta + E_y^{in}\sin\theta$ and $-E_x^{in}\sin\theta + E_y^{in}\cos\theta$, respectively. This rotation will indeed impact the $E_x$ field and can be utilized to decrease the maximal of it, thereby enhancing the contrast for interference.

**Section 4. Jones matrix metasurface design with nanoblock elements**

The Jones matrix of metasurface is designed with $J_{xx} = 0.5e^{i\theta_x} + 0.5U_{norm}$ and $J_{yy} = 0.39e^{i\theta_y} + 0.61U_{norm}$, where the maximal amplitudes are both normalized to unity. To realize this specific Jones matrix, we employ a unit cell composed of two distinct rectangular nanoblocks labeled as $A$ and $B$. The transmission coefficients through the $x$ and $y$ axes are denoted as $\varphi_{Ax}$ and $\varphi_{Ay}$ for $A$, and $\varphi_{Bx}$ and $\varphi_{By}$ for $B$, respectively. The total Jones matrix of the unit cell can be expressed as:

$$J_{xx} = A_{xx}e^{i\delta_{xx}} = \frac{1}{2}(e^{i\varphi_{Ax}} + e^{i\varphi_{Bx}}) \tag{S21}$$

$$J_{yy} = A_{yy}e^{i\delta_{yy}} = \frac{1}{2}(e^{i\varphi_{Ay}} + e^{i\varphi_{By}}) \tag{S22}$$

Here, the factor 1/2 is a normalized coefficient for the Jones matrix. The four phase terms $\varphi_{Ax}$, $\varphi_{Ay}$, $\varphi_{Bx}$ and $\varphi_{By}$ can be directly obtained as:

$$\varphi_{Ax} = \delta_{xx} + \text{acos}(A_{xx}) \tag{S23}$$

$$\varphi_{Bx} = \delta_{xx} - \text{acos}(A_{xx}) \tag{S24}$$

$$\varphi_{Ay} = \delta_{yy} + \text{acos}(A_{yy}) \tag{S25}$$

$$\varphi_{By} = \delta_{yy} - \text{acos}(A_{yy}) \tag{S26}$$

For the nanoblock structure, it supports two propagation modes along its $x$ and $y$ axes. The transmission magnitude and phase shift with $x$-polarized and $y$-polarized incidences, as functions of the transverse dimensions of the nanoblocks dx and dy are illustrated in supplementary fig. 4. The period is 350 nm, the height of the nanoblock is 600 nm and the wavelength is 780 nm. To achieve a desired phase retardation of $\varphi_x$ and $\varphi_y$, the transverse dimensions of the nanoblocks are selected according to the following steps:

(1) Begin by setting a pre-defined average transmission magnitude $t_{avg}$. This value is primarily determined by the overall transmissions of the nanoblocks with different transverse dimensions and should not be too small. In our work, we have chosen $t_{avg} = 0.98$.

(2) Calculate the complex-valued errors $\varepsilon_x = \left| t_{avg} e^{i\varphi_x} - t_{simulated} e^{i\varphi_{x,simulated}} \right|$ and $\varepsilon_x = \left| t_{avg} e^{i\varphi_x} - t_{simulated} e^{i\varphi_{x,simulated}} \right|$, and choose the large one $\varepsilon_{max} = \max(\varepsilon_x, \varepsilon_y)$.

(3) Determine the configuration that minimizes $\varepsilon_{max}$ for all possible dimensions.

These steps help in obtaining the desired phase shifts by selecting appropriate transverse dimensions for the nanoblocks while maintaining an acceptable transmission magnitude.

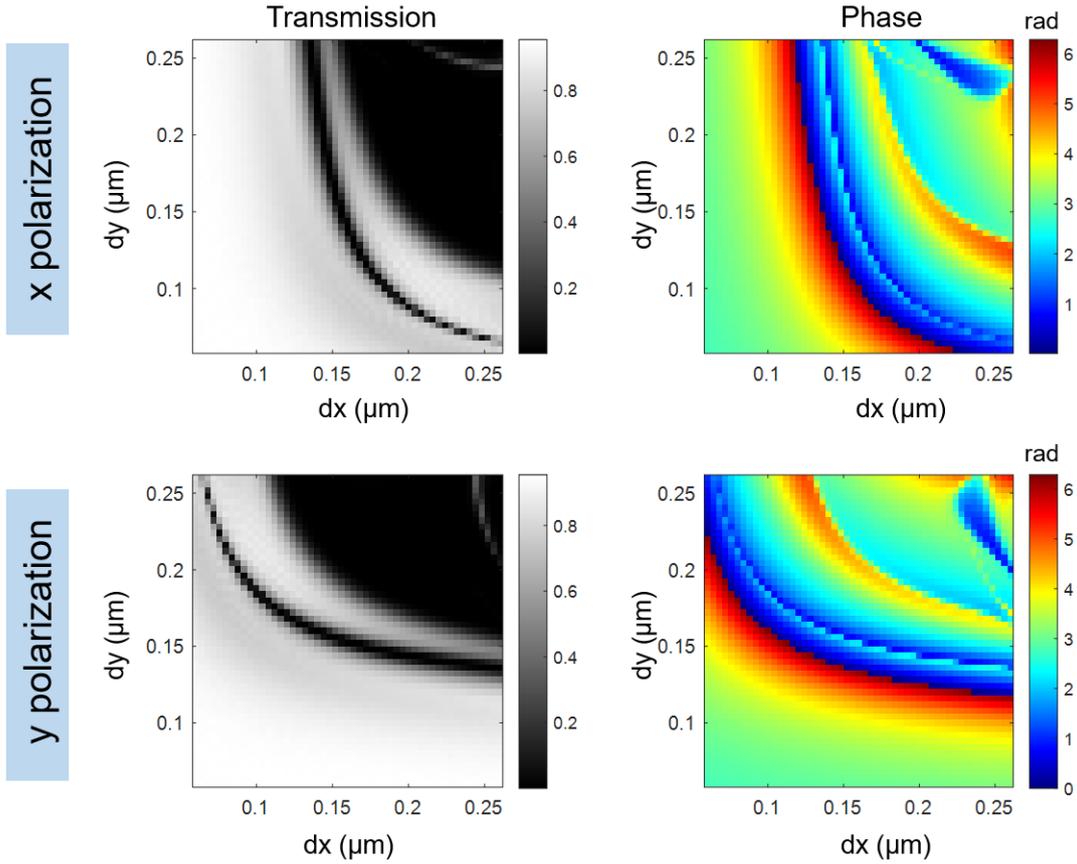

Supplementary Figure 4. Metasurface unit design. Transmission and phase shift (radians) of the nanoblock as a function of the transverse dimensions, dx and dy, for *x*-polarized and y-polarized incidences. The optical response for *y*-polarization response is obtained by swapping *x* and *y* of x-polarization response due to the nature of symmetry.

**Section 5. Encoding arbitrary complex-amplitude distribution with pure-phase SLM**

In this section, we aim to use the Spatial Light Modulator (SLM) to impose arbitrary amplitude and phase distributions with pure phase modulation. The SLM used in our setup is the PLUTO-2-NIR-011 from HOLOEYE, which is a phase-only SLM with a pixel pitch of $P$=8 μm. At the designed wavelength of 780 nm, the SLM enables an effective diffraction numerical aperture (N.A. sin$\theta$) of $\lambda/2P$=0.0488 with the center located at zero.

On the other hand, the designed metasurface has a diameter of 600 μm, and the focal length of the lens in the 4$f$ system, as shown in Fig. 2a, is 15 mm, indicating an N.A. of 0.02. It might appear that the diffraction field generated by the SLM can cover the required N.A. for the metasurface. However, this is not the case. When using pure phase modulation to generate a complex amplitude, the N.A. of this complex-amplitude field is inherently limited to a very small value, much smaller than the N.A. of the pure phase distribution.

To overcome this limitation, we employ two lenses to scale the SLM patterns. In Fig. 3c, the focal lengths of lenses 1 and 2 are chosen as 150 mm and 50 mm, respectively. As a result, the generated arbitrary field has a period of 8/3=2.67 μm and the effective diffraction N.A. is increased to 0.14625. With such an N.A. of pure phase, we can generate arbitrary complex amplitudes within an N.A. of 0.02.

Assuming that the generated arbitrary complex amplitude is represented as $U(x,y)=A(x,y)\exp(1i\varphi(x,y))$, with a normalized $A$ and the center spatial spectrum located at zero, the pixel number is set to 360×360, ensuring its size remains less than 1 mm. We have chosen three different optical field configurations, as shown in the first column in supplementary fig. 5. The phase distribution for the three types is as follows:

For $E_x$ field (type 1):

$\varphi(x, y) = 3\theta$ (where $\theta$ is the azimuthal angle) \hfill (S27)

For $E_y$ field (type 1):

$$\varphi(x, y) = \frac{2\pi}{750\mu m} r - 3.0 \text{ (where r is the azimuth radius)} \tag{S28}$$

For $E_x$ field (type 2):

$$\varphi(x, y) = -5\left(\frac{x}{400\mu m}\right)^2 - 3\left|\frac{y}{400\mu m}\right| \tag{S29}$$

For $E_y$ field (type 2):

$$\varphi(x, y) = 3\left(\frac{r}{400\mu m}\right)^2 + 3\left|\frac{r}{400\mu m}\right| + 3\cos\left(\frac{x}{150\mu m}\right) - 0.4 \tag{S30}$$

For $E_x$ field (type 3):

$$\varphi(x, y) = 3\sin\left(\frac{x}{250\mu m} + 1\right)^2 + 3\cos\left(\frac{y}{200\mu m}\right) \tag{S31}$$

For $E_y$ field (type 3):

$$\varphi(x, y) = 3.5\sin\left(\frac{x+y}{250\mu m}\right) + 1.8\left(\frac{x}{400\mu m}\right)^2 - 0.4 \tag{S32}$$

Firstly, we limit the N.A. of these fields to 0.02 using a Gaussian filter and spectrum truncation. The results are shown in the second column in supplementary fig. 5, which exhibit no significant difference from the original fields but with a slight blur.

Subsequently, we employ type 3 of the pure phase holograms [1] to generate the desired complex amplitude. In detail, we first add a phase term $-k\delta x$ to $\varphi(x,y)$, i.e., $\varphi'(x,y) = \varphi(x,y) - k\delta x$, where $k = 2\pi/\lambda$ and $\delta = 0.123$. This adjustment shifts the center of the N.A. of $U(x,y)$ to $(-\delta, 0)$. We can then obtain the pure phase of the SLM, denoted as $\Psi = \text{inversebessel1}(0.5819A)\sin(\varphi')$, where inversebessel1 is the inverse of the first-order Bessel function. This pure phase distribution generates the desired complex field with an N.A. of 0.02 at center of ($-\delta$, 0). We subsequently shift the center to zero by adding a phase term of $k\delta x$, i.e., $\exp(ik\delta x)\cdot\exp(i\Psi)$, with the phase shown in the third column of supplementary fig. 5. With this pure phase distribution, the obtained amplitude and phase with limited N.A. of 0.02 (fourth column in supplementary fig. 5) closely align with our design.

Finally, we apply this phase to the SLM. It's important to note that a gradient phase term needs to be added to compensate for the gradient phase introduced by oblique incidence. The fifth column in supplementary fig. 5 presents the measured intensity of the field generated using the pure phase on the SLM. It closely aligns with our design, albeit with some speckle noises present. When the metasurface is positioned at the center of the Fourier plane, the generated sub-images exhibit similar speckle noise patterns. This observation suggests that the metasurface has effectively diffracted the generated images into multiple orders, and the speckle noise is not a consequence of the metasurface itself. The obtained images with (Fig. 4 in the main text) and without the metasurface (fifth column in supplementary fig. 5) are compared, with structural similarity (SSIM) values around 0.82.

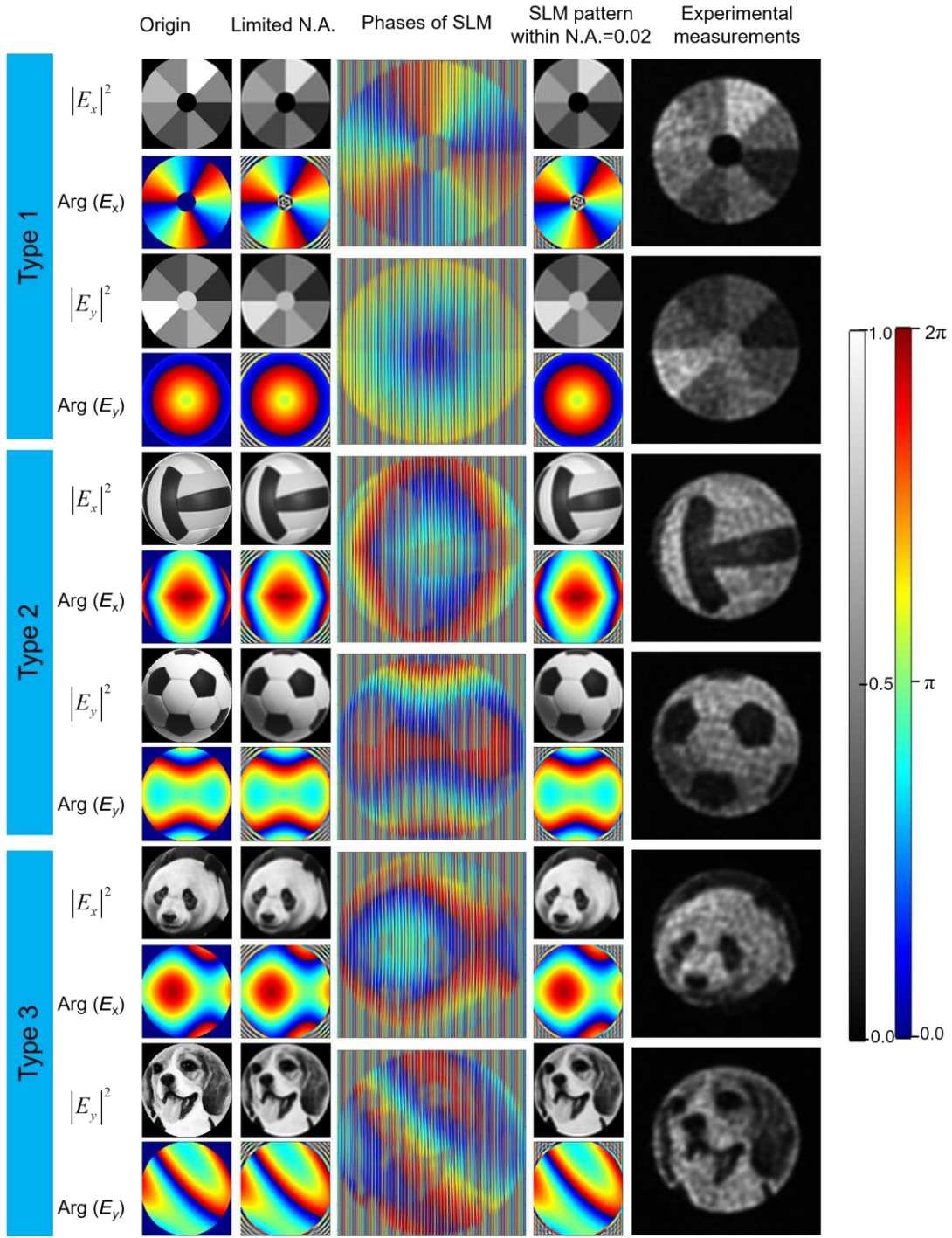

Supplementary Figure 5. Encoding three types of complex-amplitude distribution with pure-phase modulation. The figure consists of five columns. First column displays the original amplitude and phase distributions. Second column depicts the amplitude and phase distributions limited by N.A.=0.02 through the use of a Gaussian filter and spectrum truncation. Third column shows the pure phase distribution employed to generate the corresponding complex amplitude. Fourth column illustrates the complex amplitude resulting from the pure phase distribution within N.A.=0.02. Fifth column shows the experimental measurement of intensity distribution with pure phase imposed on the SLM.

**Section 6. Lens choose and image correction**

Recalling the optical system based on the metasurface (Figure 2b of main text), it consists of two lenses and a metasurface. The input field is positioned at the center of the front focal plane of Lens 1 and is focused at the center of the back focal plane. Consequently, for Lens 1, a conventional spherical lens is a suitable choice, as it introduces minimal spherical aberration. However, as the field passes through the metasurface, it undergoes diffraction into several orders with large diffraction angles and is subsequently imaged at regions located at a certain distance from the center. This results in significant spherical aberration. To mitigate this spherical aberration, we select an aspheric lens (AL1815, Thorlabs, Inc.) for Lens 2.

We conduct a ray tracing simulation to compare two scenarios: one with a conventional spherical lens and the other with an aspheric lens for Lens 2. The results are presented in supplementary fig. 6. In this simulation, a grating is placed at the center Fourier plane to diffract the incident field with $\sin(\theta_x)=2/15$ and $\sin(\theta_y)=-2/15$, corresponding to the one at the corner of the seven sub-images. It can be observed that the aspheric lens can achieve better focus with clearer imaging compared to the spherical lens, thereby reducing spherical aberration. However, both cases exhibit some degree of image distortion due to the large diffraction angle.

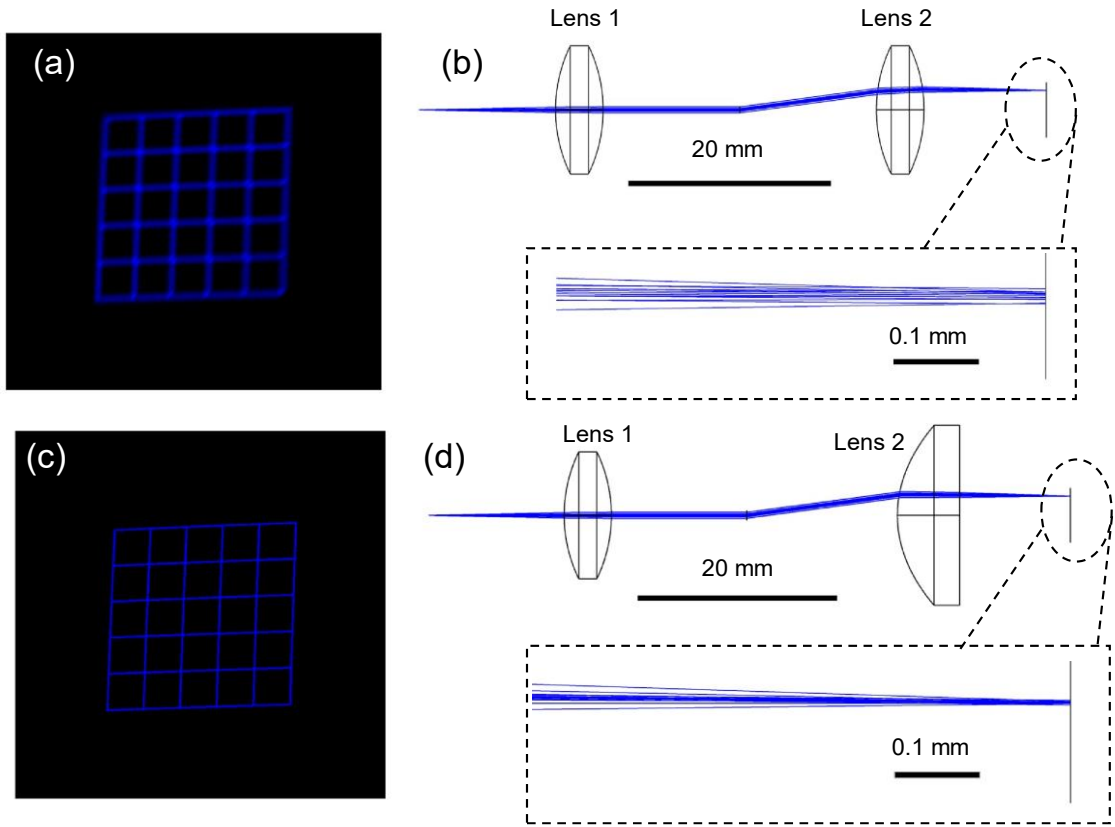

Supplementary Figure 6. Ray tracing simulation comparison between two different types of lenses. (a-b) Image simulation of a grid pattern (a) and point-to-point imaging (b) with a spherical lens for Lens 2. (c-d) Image simulation of a grid pattern (c) and point-to-point imaging (d) with an aspheric lens for Lens 2. The enlarged insets in (b) and (d) provide a closer look, highlighting the superior focusing capabilities of the optical system employing the aspheric lens.

To correct this distortion, a checkerboard image (supplementary fig. 7a) is employed as the input, and its focused images of the seven diffraction orders are simulated by implementing gratings with different diffraction angles. The results are shown in supplementary fig. 7b, where various levels of distortion can be observed in all seven images. In our experiment, we observe similar distortions, as demonstrated in the center column of Figure 4 in the main text.

Subsequently, the pattern points of the checkerboard are detected and labeled with circles in supplementary figs. 7a-b. The points in supplementary fig. 7a serve as the base control points, and the points in the seven sub-images (supplementary fig. 7b) are used as input points to be transformed. We utilize the "cp2tform" function in MATLAB to compute a transformation that best fits the mapping from the input points to the base points, employing a polynomial transformation type, which is referred to as "tform". This transformation is then applied (using the "imtransform" function in MATLAB) to the seven distorted images, resulting in the corrected images (supplementary fig. 7c). This transformation is applied to our measured results for image correction, resulting in the full field distribution displayed in the rightmost column of Fig. 4 of main text.

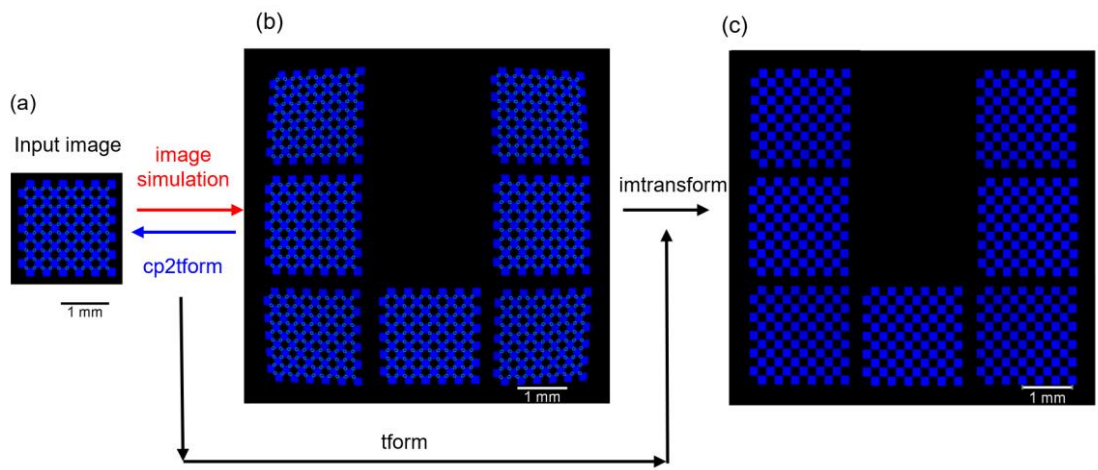

Supplementary Figure 7. Imaging correction. (a) Input checkerboard pattern. (b) Simulated images with ray tracing for the seven diffraction orders. The pattern points in (a) and (b) are marked with circles and are used as reference control points and input

points for transformation correction. The correction is achieved using the "cp2tform" function in Matlab. (c) The transformed images of the seven orders with the "tform" obtained from the previous transformation.

**Section 7. Metasurface efficiency measurement**

To measure the efficiency of the metasurface for imaging, we initially projected the designed optical patterns onto SLM and removed the metasurface. In this configuration, the measured optical image in the CMOS camera consists of a single image, with its power measured as $I_{in}$. Then, we placed the metasurface at the center Fourier plane of the 4$f$ system and captured seven sub-images and one zero-order image in the CMOS camera. The power of the seven images of interest is measured as $I_{out}$. The efficiency is defined as $I_{out} / I_{in}$.

It's essential to note that this efficiency is not constant but depends on the input field. This is because the incident field is the Fourier transform of the input field. When the input field changes, the incident field on the metasurface changes as well, resulting in a change in the transmitted light through the metasurface. For the three specific optical fields presented in Figure 4 of the main text, the measured efficiency falls between 17.5% to 22.4%.